# Surgical Data Science: Enabling next-generation surgery

Lena Maier-Hein[1,*], Swaroop Vedula[2], Stefanie Speidel[3], Nassir Navab[4,14], Ron Kikinis[5,6,7], Adrian Park[8], Matthias Eisenmann[1], Hubertus Feussner[9], Germain Forestier[10], Stamatia Giannarou[11], Makoto Hashizume[12], Darko Katic[3], Hannes Kenngott[13], Michael Kranzfelder[9], Anand Malpani[2,14], Keno März[1], Thomas Neumuth[15], Nicolas Padoy[16], Carla Pugh[17], Nicolai Schoch[18], Danail Stoyanov[19], Russell Taylor[14], Martin Wagner[10], Gregory D. Hager[2,14,x], Pierre Jannin[20,21,x]

[1] Department of Computer Assisted Medical Interventions (CAMI), German Cancer Research Center (DKFZ), Heidelberg, Germany

[2] The Malone Center for Engineering in Healthcare, The Johns Hopkins University, Baltimore, Maryland, USA

[3] Institute for Anthropomatics and Robotics, Karlsruhe Institute of Technology, Karlsruhe, Germany

[4] Chair for Computer Aided Medical Procedures, Technical University of Munich, Munich, Germany

[5] Department of Radiology, Brigham and Women's Hospital and Harvard Medical School, Boston, MA, USA

[6] Department of Computer Science, University of Bremen, Bremen, Germany

[7] Fraunhofer MEVIS, Bremen, Germany

[8] Department of Surgery, Center of Anne Arundel Health System, Annapolis, MD, USA

[9] Department of Surgery, Klinikum rechts der Isar, Technical University of Munich, Munich, Germany

[10] Department of Computer Science, University of Haute-Alsace, Mulhouse, France

[11] The Hamlyn Centre for Robotic Surgery, Imperial College London, UK

[12] Department of Advanced Medical Initiatives, Graduate School of Medical Sciences, Kyushu University, Fukuoka, Japan

[13] Department for General, Visceral and Transplant Surgery, Heidelberg University Hospital, Heidelberg, Germany

[14] Department of Computer Science, The Johns Hopkins University, Baltimore, MD, USA

[15] Innovation Center Computer Assisted Surgery (ICCAS), University of Leipzig, Leipzig, Germany

[16] ICube, University of Strasbourg, CNRS, IHU Strasbourg, France

[17] Department of Surgery, University of Wisconsin, Madison, WI, USA

[18] Engineering Mathematics and Computing Lab (EMCL), IWR, Heidelberg University, Heidelberg, Germany

[19] Centre for Medical Image Computing (CMIC) & Department of Computer Science, University College London, London, UK

[20] Université de Rennes 1, Rennes, France

[21] INSERM, Rennes, France

[*] Please send correspondence to: l.maier-hein@dkfz.de

[x] Shared senior authors

**Future advances in surgical care require, to a greater and greater extent, a close partnership between caregivers, patients, technology, and information systems. As part of the move towards personalized medicine, interventional care will increasingly transform from an artisanal craft based on physicians' individual experiences, preferences and traditions into a discipline that relies on objective decision-making based on large-scale data from heterogeneous sources.**

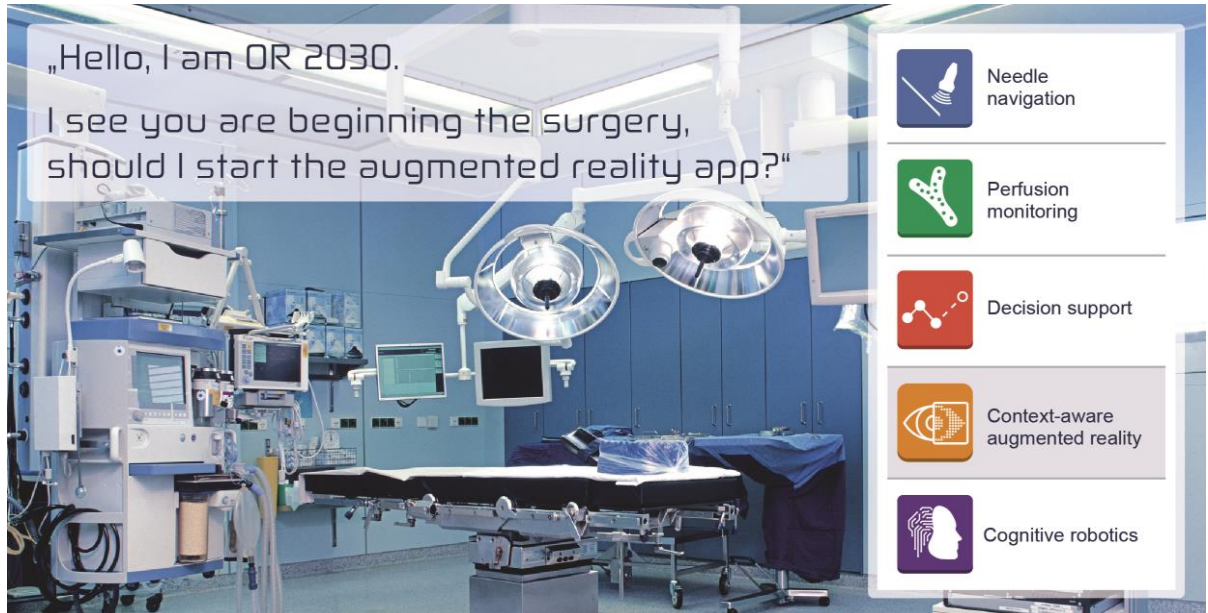

Figure 1: OR 2030. The operating room of the future will be seamlessly synchronized with the surgical procedure to provide the right assistance at the right time.

Data science is an emerging interdisciplinary field that deals with the extraction of knowledge from data. Despite the tremendous progress in the field of data science made over the past decade, there has been a delay in introducing large-scale data science into interventional medicine (e.g. surgery, interventional radiology, gastroenterology, radiotherapy). This delay can partly be attributed to the fact that, today, only a fraction of patient-related data and information is digitized and stored in a structured and standardized manner, e.g. in registries (1,2). Furthermore, diversity in caregiver training, experience and routine institutional practices have elicited variation in methods and means of perioperative care. Without data to provide an insight into actual practice, disparity in outcomes is an inevitable consequence.

This paper introduces Surgical Data Science as an emerging scientific discipline. Key perspectives emerged, based on discussions held in an intensive two-day international interactive workshop[1] that brought together leading researchers working in the related field of computer and robot assisted interventions. Our consensual opinion is that increasing access to large amounts of complex data throughout the patient care process, complemented by advances in data science and machine learning techniques, has set the stage for a new generation of analytics that will support decision-making and quality improvement in interventional medicine. In this article, we provide a consensual definition for Surgical Data Science, identify associated challenges and opportunities and provide a roadmap for advancing the field.

---

[1] www.surgical-data-science.org/workshop2016

# Evolution of Surgical Practice

"Surgery is a profession defined by its authority to cure by means of bodily invasion" (3). Despite increased expectation for outcomes and safety from patients, hospitals and insurers, studies estimate that 9 million (4) of an estimated 300 million surgical procedures per year worldwide (5) will encounter major complications.

Surgical practice has significantly evolved throughout the centuries (cf. Fig. 2). It underwent revolutionary changes with the introduction of anesthesia and antiseptics in the 19$^{th}$ century. At this point, surgeons typically relied on minimal instrumentation as well as their own knowledge and clinical experience, which was, to some extent, augmented by learning from peers and few available medical books. In the 20$^{th}$ century, advances in surgery centered around professionalization, systematic measurement of outcomes of care, and minimally invasive access to surgical sites. Surgery was further transformed with the introduction of multimodal medical imaging (6), the development of surgical microscopes and endoscopes and ultimately the emergence of computer and robot assisted interventions (7). Despite rapid advances, the seamless integration of computer-aids in the surgical environment which enhance situation awareness, ergonomics and minimization of cognitive workload has not yet been achieved. Furthermore, the internet revolution has brought access to an almost unlimited amount of electronic patient records, but this avalanche of data is typically unstructured with limited quality control and almost no direct integration with computer-assisted surgical systems.

Future advances in surgery will continue to be motivated by safety, effectiveness, and efficiency of care. The next paradigm shift will be from implicit to explicit models, from subjective to objective decision-making, and from qualitative to quantitative assessment. This will enable personalized treatment and will ensure that future evolution is centred around patients and caregivers. Within this vision of the future, Surgical Data Science will evolve to observe everything occurring within and around the treatment process. It will provide the surgeon with quantitative support to aid decision-making and surgical actions and - importantly - will link decisions to patient outcomes. For the patient, this will mean having access to the best surgical care with less variability arising from unique patient characteristics rather than the choice of surgeon or care facility. Ultimately, Surgical Data Science will offer the opportunity to create "superhuman" surgery by moving beyond the data associations that individuals are able to perceive, detect and maintain, into the realm of vast data types and sizes that can only be exploited through modern computing solutions.

# What is Surgical Data Science?

While Surgical Data Science is related to the field of Biomedical Data Science, its unique characteristic is the focus on *procedural* data. It pertains to (i) the patient, (ii) effectors involved in the manipulation of the patient including physicians, anesthesia team, nurses and devices, including robots, (iii) sensors for perceiving patient- and procedure-related data such as images, vital signs, medical device data and motion data as well as (iv) domain knowledge, including *factual knowledge*, such as (hospital-specific) standards related to the clinical workflow, previous findings from studies or clinical guidelines as well as *practical knowledge* from previous procedures.

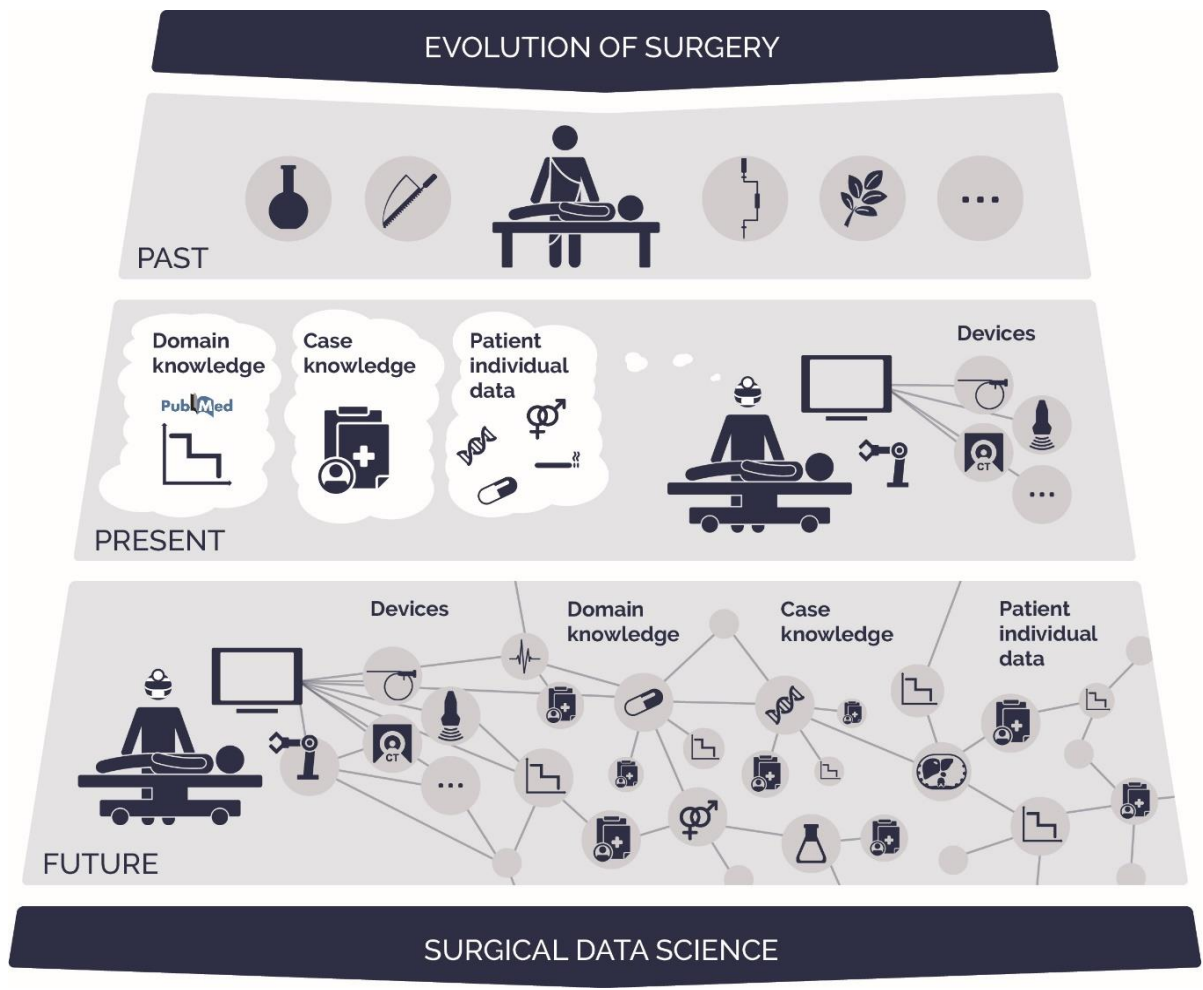

Figure 2: Evolution of Surgery: In the PAST, a "physician for all purposes" handled patient treatment based on local traditions with only a minimum of equipment. At PRESENT, a wealth of information can be acquired for each patient, and modern surgery rooms are equipped with numerous devices for performing and monitoring treatment. However, it is up to the individual surgical team to make use of their domain knowledge and experience to use all the available information in an optimal manner. FUTURE surgery will be based on automatic holistic processing of all the available data to facilitate, optimize and objectify care delivery using Surgical Data Science techniques.

Consensual definition: *Surgical Data Science is an emerging scientific field with the objective of improving the quality of interventional healthcare and its value through capturing, organization, analysis, and modeling of data. It encompasses all clinical disciplines in which patient care requires intervention to manipulate anatomical structures with a diagnostic, prognostic, or therapeutic goal, such as surgery, interventional radiology, radiotherapy, and interventional gastroenterology. Data may pertain to any part of the patient care process (from initial presentation to long-term outcomes), may concern the patient, caregivers, and/or technology used to deliver care, and is analyzed in the context of generic domain-specific knowledge derived from existing evidence, clinical guidelines, current practice patterns, caregiver experience, and patient preferences. Data may be obtained through medical records, imaging, medical devices or sensors that may either be positioned on patients or caregivers or integrated into the instruments and technology used to deliver care. Improvement may result from understanding processes and strategies, predicting events and clinical outcome, assisting physicians in decision-making and planning execution, optimizing ergonomics of systems, controlling devices before, during and after treatment as well as from advances in prevention, training, simulation and assessment. Surgical Data Science builds on principles and methods from other data-intensive disciplines such as computer science,*

*engineering, information theory, statistics, mathematics, and epidemiology, and complements other information-enabled technologies such as surgical robotics, smart operating rooms, and electronic patient records.*

## Key Clinical Applications

As the definition above suggests, a data science approach may impact interventional care throughout the entire patient care pathway. Some of the opportunities for impact in the specific context of surgery include:

i) *Decision Support*

The quality of surgical care is affected to varying extents by decisions made by caregivers and patients throughout the care pathway. Traditionally, surgeons relied upon their experience to play a major role in consequential decisions such as whether to operate and the type of surgery to be performed (8). This decision-making model has gradually evolved to be informed by predictive analytics based on systematic data capture and curation through patient registries. However, currently available registry-based analytics to support surgical decision-making rely upon cross-sectional measures of a subset of patient characteristics before surgery (9). Furthermore, registries rarely capture the full record of the patient care pathway and the amount of data that they are missing varies (10). A data science approach to decision-support relies not only upon continuously updating predictive analytics throughout the patient care process but also upon more comprehensive and unconventional sources of data (11,12,13). Furthermore, surgical decisions may be optimized by modeling individual patients within the context of population-level data and other multimodal data sources (14,15). Finally, Surgical Data Science reinforces the importance of integration of such decision-support into patient care workflows via user-friendly data products.

ii) *Context-aware Assistance*

Surgical Data Science enables context-aware assistance and can be applied throughout the patient care pathway. In the operating room, application can include monitoring procedures to predict remaining duration to facilitate scheduling or to anticipate needs for resources (16). Similarly, autonomous assistance can provide surgeons with timely information through surgical phase recognition (17,18), decision-support through patient-specific simulations (19), and collaborative robots (20). Context-aware assistance improves the safety, quality, and efficiency of care and can augment providers' performance when integrated into surgical care pathways.

iii) *Surgical Training*

Surgical education and certification ensure that competent surgeons provide care, and are thus a critical element in assuring quality of care. Poor surgical technical skill is associated with an increased risk of readmission, reoperation, and death (21,22). Technical skills and errors are also associated with non-technical skills such as decision-making (23). Surgical Data Science can be transformative for surgical training through objective computer-aided skill evaluation (OCASE) (24), robot-assisted active learning of technical skills (25), patient- and context-specific simulation training and assessment, and surgical coaching (26,27). Additional data analytics such as surgical process modeling, detection of constituent activities, errors, and skill deficits facilitate targeted feedback based on OCASE (28,29).

Surgical Data Science thus represents the new frontier for surgical training in a complex patient care environment with limited resources.

## Key Challenges

We foresee two immediate challenges to advancing our vision of Surgical Data Science - data availability and analysis of highly heterogeneous multi-modal data.

Surgical Data Science relies upon access to high-quality data on a large scale that documents both the patient care process and patient outcomes. While other communities share databases for advancing research and practice (cf. e.g. ImageNet[2]), such resourceful databases are lacking in documentation of surgery despite it being inherent that quality improvement can be achieved through outcome measurement, for example, using patient registries. This paucity of databases may be attributed to a multitude of regulatory, technical, and sociological factors. For example, concerns related to privacy and confidentiality of both patients and caregivers pose important legal and ethical issues that must be addressed for data science to be possible. On the other hand, although large amounts of data are routinely available during interventional care, they are not captured and annotated using standardized protocols (30). While international healthcare terminology standards for biomedical data science are well-established (cf. e.g. Foundational Model of Anatomy (FMA)[3], Gene Ontology (GO)[4], SNOMED-CT[5]), ontologies to describe activities and other aspects of interventional care processes are lacking. Furthermore, data annotation is resource-intensive. Although some aspects of data annotation for interventional care processes can be crowdsourced to untrained lay individuals (31), others may require expertise for their content. Ultimately, data should be collected as a matter of best practice in a consistent, longitudinal manner using tools that smoothly integrate into the clinical workflow. Workers in the field need to identify allies and clear short-term "win scenarios" that will build interest and trust in the area so that hospitals, insurers, and practitioners all see the value of creating these resources which will ultimately advance the profession (32).

Analysis of data from interventions also introduces unique challenges. Firstly, a substantial aspect of Surgical Data Science involves modeling the orchestrated manipulation by teams of individuals and patients' response to such actions. In surgical procedures, for example, not only the head surgeon but also anesthetists, assistant surgeons, circulators and nurses, play crucial roles at different workflow steps within surgery and their smooth dynamic collaboration and coordination play an important role in the success of the overall process. Second, anatomical manipulation during surgery is frequently irreversible, with errors resulting in serious complications or even death. Therefore the robustness and reliability of the methods are of crucial importance (33). Furthermore, while the diagnostic process follows a rather regular flow of data acquisition and big companies such as Google Inc. (Mountain View, CA, USA) and IBM (Armonk, NY, United States) have started developing Biomedical Data Science techniques to support it, the surgical process varies significantly from case to case and is highly specific to procedure, patient, and surgeon (34). The heterogeneity in the data resulting from different hardware, imaging protocols (cf. OR.NET[6] and MD PnP[7]), context, training, care guidelines, physicians, and so forth is a great challenge to be overcome - not only for the development of data analysis methods but also for the validation of new methodology and systems. Finally, procedural data must be holistically analyzed with other heterogeneous data including genetics, biomarkers, patient demographics, imaging and pre- and intraoperative data, enabling us to move from eminence-based to knowledge-based and data-driven medicine. In this context, shared tools for optimizing discovery and training researchers could significantly advance the field (35).

---

[2] www.image-net.org
[3] www.si.washington.edu/projects/fma
[4] www.geneontology.org/
[5] www.snomed.org/
[6] www.ornet.org

## Dissemination and Impact

Surgical Data Science is a field of scientific research. It enables fundamental understanding of surgical procedures, their variability, crucial parameters, hidden structures, dependencies, optimal pathways, the importance of each parameter and keys to success and failure of methodologies and the basic principles driving our surgical education, training and practice. In this sense, its dissemination will be manifold. As discussed above, Surgical Data Science could change the education and training of millions of physicians across the planet. Wikipedia has allowed us to accumulate, prune and improve our knowledge and make it available to billions; and in the same way, search engines allow us to access information instantaneously and receive information with ease. Similarly, Surgical Data Science will facilitate the methods via which the next generation of medical students learn from complex data without restricting them to a particular book or a particular teacher. We expect that distinct career pathways will evolve for training Surgical Data Scientists and embedding them into clinical research teams. In addition, data science may be introduced into undergraduate and medical school curricula.

The end-point for discoveries through Surgical Data Science is their effective translation into patient care workflows, which can involve commercialization of data products and services. This is possible when various stakeholders, such as academic scientists and commercial partners, collaborate from inception through to translation of data products. Surgical Data Science offers a diverse space for discovery and innovation, which may transform into a wide range of products such as decision support systems, smart instrumentation, intelligent technologies, or surgical training. Surgical Data Science will enable medical companies to fully optimize all of their solutions and also allow in-depth usability studies of each component of every surgical product based on large amount of data and its interaction with all the other components and players in this complex domain.

In summary, Surgical Data Science can be disseminated through its impact on a wide range of products, from medical training and education to surgical imaging, instrumentation and user interface, and next-generation advanced patient information systems can also be constantly updated based on analysis of large amounts of dynamic data.

---

[7] www.mdpnp.org

| **Towards next-generation surgery** |
|---|
| ● Surgical Data Science will pave the way from artisanal to data-driven interventional healthcare with concomitant improvements in quality and efficiency of care.<br>● A key element will be to institutionalize a culture of continuous measurement, assessment and improvement using evidence from data as a core component.<br>● An actionable path would be that societies support and nurture efforts in this direction through best practice, comprehensive data registries, and active engagement and oversight.<br>● Surgical Data Science should be established as a new element of both the education and career pathway for hospitals that teach and train future interventionalists. |

## Acknowledgements

The authors wish to thank the Collaborative Research Center (SFB/TRR) 125: Cognition-guided Surgery, funded by the German Research Foundation (DFG), for sponsoring the workshop that served as basis for the manuscript ([www.surgical-data-science.org/workshop2016)](www.surgical-data-science.org/workshop2016)). Many thanks also go to all workshop participants for their valuable input during the workshop and to Carolin Feldmann for preparing the figures.

## References

1. Z. Obermeyer, E. J. Emanuel, Predicting the Future — Big data, machine learning, and clinical medicine, *New England Journal of Medicine* **375**, 1216–1219 (2016).

2. M. E. Porter, S. Larsson, T. H. Lee, Standardizing patient outcomes measurement, *New England* Journal *of Medicine* **374**, 504–506 (2016).

3. A. Gawande, Two hundred years of surgery, *New England Journal of Medicine* **366**, 1716–1723 (2012).

4. T. G. Weiser, S. E. Regenbogen, K. D. Thompson, A. B. Haynes, S. R. Lipsitz, W. R. Berry, A. A. Gawande, An estimation of the global volume of surgery: a modelling strategy based on available data, *The Lancet* **372**, 139–144 (2008).

5. T. G. Weiser, A. B. Haynes, G. Molina, S. R. Lipsitz, M. M. Esquivel, T. Uribe-Leitz, R. Fu, T. Azad, T. E. Chao, W. R. Berry, A. A. Gawande, Estimate of the global volume of surgery in 2012: an assessment supporting improved health outcomes, *The Lancet* **385**, **Supplement 2**, S11 (2015).

6. Z. H. Cho, J. P. Jones, M. Singh, *Foundations of Medical Imaging* (Wiley, New York, 1993).

7. K. Cleary, T. M. Peters, Image-guided Interventions: Technology review and clinical applications, *Annual Review of Biomedical Engineering* **12**, 119–142 (2010).

8. L. G. Glance, T. M. Osler, M. D. Neuman, Redesigning surgical decision making for high-risk patients, *New England Journal of Medicine* **370**, 1379–1381 (2014).

9. T. O. Mitchell, J. L. Holihan, E. P. Askenasy, J. A. Greenberg, J. N. Keith, R. G. *Martindale*, J. S. Roth, M. K. Liang, Do risk calculators accurately predict surgical site occurrences?, *Journal of Surgical Research* **203**, 56–63 (2016).

10. H. Lyu, M. Cooper, K. Patel, M. Daniel, M. A. Makary, Prevalence and data transparency of national clinical registries in the United States, *Journal for Healthcare Quality* **38**, 223–234 (2016).

11. P. C. Sanger, G. H. van Ramshorst, E. Mercan, S. Huang, A. L. Hartzler, C. A. L. Armstrong, R. J. Lordon, W. B. Lober, H. L. Evans, A prognostic model of surgical site infection Using Daily Clinical Wound Assessment, *Journal of the American College of Surgeons* **223**, 259–270.e2 (2016).

12. C. Ke, Y. Jin, H. Evans, B. Lober, X. Qian, J. Liu, S. Huang, Prognostics of surgical site infections using dynamic health data, *Journal of Biomedical Informatics* **65**, 22–33 (2017).

13. F. Lalys, C. Haegelen, M. Mehri, S. Drapier, M. Vérin, P. Jannin, Anatomo-clinical atlases correlate clinical data and electrode contact coordinates: Application to subthalamic deep brain stimulation, *Journal of Neuroscience Methods* **212**, 297–307 (2013).

14. K. E. Henry, D. N. Hager, P. J. Pronovost, S. Saria, A targeted real-time early warning score (TREWScore) for septic shock, *Science Translational Medicine* **7**, 299ra122 (2015).

15. K. März, M. Hafezi, T. Weller, A. Saffari, M. Nolden, N. Fard, A. Majlesara, S. Zelzer, M. *Maleshkova*, M. Volovyk, N. Gharabaghi, M. Wagner, G. Emami, S. Engelhardt, A. Fetzer, H. Kenngott, N. Rezai, A. Rettinger, R. Studer, A. Mehrabi, L. Maier-Hein, Toward knowledge-based liver surgery: holistic information processing for surgical decision support, *International Journal of Computer Assisted Radiology and Surgery* **10**, 749–759 (2015).

16. S. Franke, J. Meixensberger, T. Neumuth, Intervention time prediction from surgical low-level tasks, *Journal of Biomedical Informatics* **46**, 152–159 (2013).

17. N. Padoy, T. Blum, S. A. Ahmadi, H. Feussner, M. O. Berger, N. Navab, Statistical modeling and recognition of surgical workflow, *Medical Image Analysis* **16**, 632–641 (2012).

18. D. Katić, J. Schuck, A. L. Wekerle, H. Kenngott, B. P. Müller-Stich, R. Dillmann, S. Speidel, Bridging the gap between formal and experience-based knowledge for context-aware laparoscopy, *International Journal of Computer Assisted Radiology and Surgery* **11**, 881–888 (2016).

19. N. Schoch, F. Kißler, M. Stoll, S. Engelhardt, R. de Simone, I. Wolf, R. Bendl, V. Heuveline, Comprehensive patient-specific information preprocessing for cardiac surgery simulations, *International Journal of Computer Assisted Radiology and Surgery* **11**, 1051–1059 (2016).

20. A. Shademan, R. S. Decker, J. D. Opfermann, S. Leonard, A. Krieger, P. C. W. Kim, Supervised autonomous robotic soft tissue surgery, *Science Translational Medicine* **8**, 337ra64 (2016).

21. M. Nathan, J. M. Karamichalis, H. Liu, S. Emani, C. Baird, F. Pigula, S. Colan, R. R. Thiagarajan, E. A. Bacha, P. del Nido, Surgical technical performance scores are predictors of late mortality and unplanned reinterventions in infants after cardiac surgery, *The Journal of Thoracic and* Cardiovascular *Surgery* **144**, 1095–1101.e7 (2012).

22. J. D. Birkmeyer, J. F. Finks, A. O'Reilly, M. Oerline, A. M. Carlin, A. R. Nunn, J. Dimick, M. Banerjee, N. J. O. Birkmeyer, Surgical skill and complication rates after bariatric surgery, *New England Journal of Medicine* **369**, 1434–1442 (2013).

23. J. N. Nathwani, R. M. Fiers, R. D. Ray, A. K. Witt, K. E. Law, S. DiMarco, C. M. Pugh, Relationship between technical errors and decision-making skills in the junior resident, *Journal of Surgical Education* **73**, e84–e90 (2016).

24. S. S. Vedula, M. Ishii, G. D. Hager, Objective assessment of surgical technical skill and competency in the operating room, *Annual Review of Biomedical Engineering* **19**, (2017); published online (10.1146/annurev-bioeng-071516-044435).

25. Z. Chen, A. Malpani, P. Chalasani, A. Deguet, S. S. Vedula, P. Kazhanzides, R. H. Taylor, Virtual fixture assistance for needle passing and knot tying, *2016 IEEE/RSJ International Conference on Intelligent Robots and Systems (IROS)*, Daejeon, Korea, pp. 2343–2350 (2016).

26. C. C. Greenberg, H. N. Ghousseini, S. R. Pavuluri Quamme, H. L. Beasley, D. A. Wiegmann, Surgical coaching for individual performance improvement, *Annals of Surgery* **261**, 32–34 (2015).

27. P. Singh, R. Aggarwal, M. Tahir, P. H. Pucher, A. Darzi, A randomized controlled study to evaluate the role of video-based coaching in training laparoscopic skills, *Annals of Surgery* **261**, 862–869 (2015).

28. E. Rojas, J. Munoz-Gama, M. Sepúlveda, D. Capurro, Process mining in healthcare: A literature review, *Journal of Biomedical Informatics* **61**, 224–236 (2016).

29. M. Uemura, P. Jannin, M. Yamashita, M. Tomikawa, T. Akahoshi, S. Obata, R. Souzaki, S. Ieiri, M. Hashizume, Procedural surgical skill assessment in laparoscopic training environments, *International Journal of Computer Assisted Radiology and Surgery* **11**, 543–552 (2016).

30. H. U. Lemke, M. W. Vannier, The operating room and the need for an IT infrastructure and standards, *International Journal of Computer Assisted Radiology and Surgery* **1**, 117–121 (2006).

31. L. Maier-Hein, S. Mersmann, D. Kondermann, S. Bodenstedt, A. Sanchez, C. Stock, H. G. Kenngott, M. Eisenmann, S. Speidel, Can masses of non-experts train highly accurate image classifiers? A crowdsourcing approach to instrument segmentation in laparoscopic images, *17th International Conference on Medical Image Computing and Computer-Assisted Intervention – MICCAI 2014*, Boston, MA, USA, Lecture Notes in Computer Science. (Springer International Publishing, 2014), pp. 438–445 (2014).

32. E. Warren, Strengthening Research through Data Sharing, *New England Journal of Medicine* **375**, 401–403 (2016).

33. F. S. Collins, L. A. Tabak, Policy: NIH plans to enhance reproducibility, *Nature News* 505, **612** (2014).

34. F. Lalys, P. Jannin, Surgical process modelling: a review, *International Journal of Computer Assisted Radiology and Surgery* **9**, 495–511 (2014).

35. C. A. Mattmann, Computing: A vision for data science, *Nature* **493**, 473–475 (2013).